# Understanding Adsorption-Induced Structural Transitions in Metal-Organic Frameworks: From the Unit Cell to the Crystal

Carles Triguero[1], François-Xavier Coudert[1,*],
Anne Boutin[2], Alain H. Fuchs[1], Alexander V. Neimark[1,3,*]

[1]CNRS & Chimie ParisTech, Paris, France
[2]PASTEUR, ENS-UPMC-CNRS, Paris, France
[3]Department of Chemical and Biochemical Engineering,
Rutgers University, New Jersey, United States

## Abstract

Breathing transitions represent recently discovered adsorption-induced structural transformations between large-pore and narrow-pore conformations in bistable metal–organic frameworks, such as MIL-53. We present a multiscale physical mechanism of the dynamics of breathing transitions. We show that due to interplay between host framework elasticity and guest molecule adsorption, these transformations on the crystal level occur via layer-by-layer shear. We construct a simple Hamiltonian that describes the physics of host–host and host–guest interactions on the level of unit cells and reduces to one effective dimension due to the long-range elastic cell-cell interactions. We then use this Hamiltonian in Monte Carlo simulations of adsorption-desorption cycles to study how the behavior of unit cells is linked to the transition mechanism at the crystal level through three key physical parameters: the transition energy barrier, the cell-cell elastic coupling, and the system size.

* Authors to whom correspondence should be addressed. Electronic mail: aneimark@rutgers.edu, fx.coudert@chimie-paristech.fr

## I. Introduction

There has been growing interest in porous coordination polymers or metal-organic frameworks (MOFs) as a new family of nanoporous materials built from organic ligands and metal centers[1]. In particular, much attention has recently been focused on a fascinating subclass of metal-organic frameworks that behave in a remarkable stimuli-responsive fashion.[2,3] These soft porous crystals feature dynamic crystalline frameworks displaying reversible, large-amplitude structural deformations induced by various external stimuli, such as temperature, mechanical pressure, or guest adsorption. When we focus on adsorption, an intriguing physical phenomenon called "breathing" has been discovered in a subclass of flexible MOFs. Solids of the MIL-53 family[4] are prototypical materials displaying breathing transitions. This phenomenon, is displayed in abrupt changes of the framework volume triggered by adsorption of guest molecules that is explored to devise advanced adsorbents, drug delivery systems, sensors, and actuators.[5,6] This phenomenon involves a complex interplay of adsorption and elastic interactions in the solid, giving rise to structural phase transitions between what have been called large pore (lp) phase and narrow pore (np) phase.

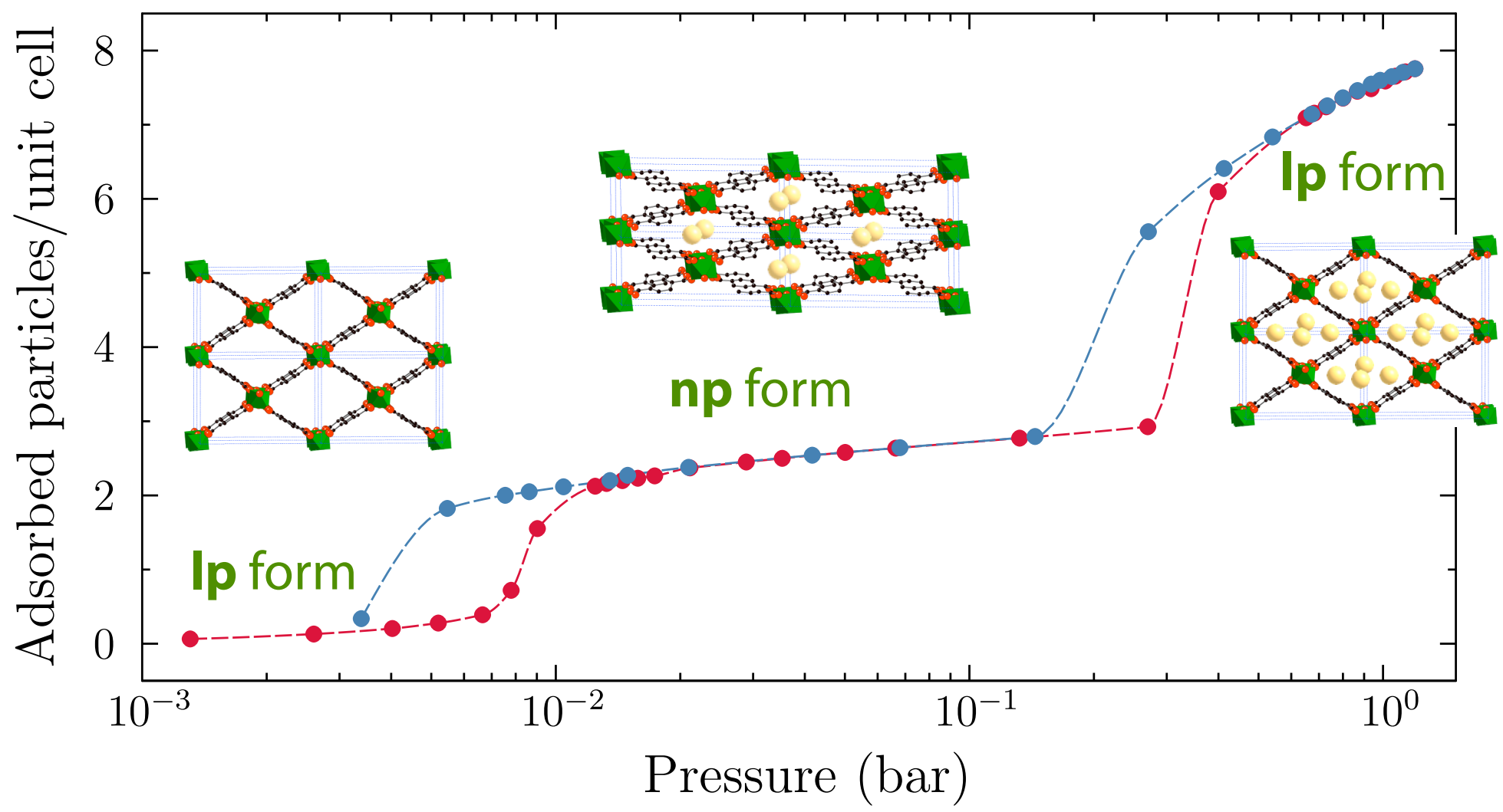

Figure 1: Adsorption (red) and desorption (blue) isotherms archetypical of the "breathing" double structural transition in materials of the MIL-53 material: experimental data for Xenon in MIL-53(Al) at 220 K.[7]

In Fig. 1 is shown the adsorption–desorption isotherm of Xenon at 220 K on a MIL-53(Al) sample,[7] which displays a typical double breathing transition. The MIL-53 framework is made of parallel one-dimensional M(OH) chains (M = $Al^{3+}$, $Cr^{3+}$, $Ga^{3+}$, ...), linked together by 1,4-benzenedicarboxylate (BDC) ligands to form linear diamond-shaped channels that are wide enough to accommodate small guest molecules. This structure may oscillate between lp and np phases, which have a remarkable difference in unit cell volume of up to 40%[8,9] (see schematics in Fig. 1). What is striking is that the transition from the larger volume lp phase to the smaller volume np phase is not necessarily accompanied with the release of guest molecules that would be expected for the normal "exhaling". For example, the equilibrium state of the MIL-53 crystal at 220 K in the absence of guest molecules is in the lp phase, and upon Xe adsorption, there first occurs the transition from the unloaded lp phase to the loaded np phase.[7] This transition is associated with a sharp uptake of Xe, from a loading of 0.2 to 2.5 molecules per unit cell, and a decrease of the crystal volume by 25%. Upon further increase of the gas pressure, adsorption gradually proceeds in the np phase up to a certain point, when the second, now "normal", breathing transiting occurs, from the np phase to the lp phase. The sample abruptly "inhales", increasing the loading from 2.7 to 6.5 molecules per cell, and expands, compensating for the volume lost upon the first lp $\rightarrow$ np transition. On the desorption pass, the reverse normal lp $\rightarrow$ np and abnormal np $\rightarrow$ lp exhaling transitions take place with a prominent hysteresis. This breathing phenomenon is engendered by guest-host adsorption interactions mediated by the elasticity of three dimensional host framework, which are currently poorly understood. The specific variations of the linker conformations in the lp and np phases during breathing transitions have been studied at the molecular level by Férey and coauthors, both experimentally (in situ X-ray diffraction)[8] and using molecular simulation (single point DFT calculations and force-field-based dynamics).[10] These works provide useful insight into the chemistry of the transformation of linker bonds associated with the framework deformation. However, a knowledge gap exists between this molecular understanding and the question of how the adsorption of guest molecules induces the physical forces responsible for macroscopic structural transformations on the sample level.

In this work, we develop a model of the dynamics of adsorption-induced deformation and structural transformation in MIL-53 type porous crystals based on the coupling of the host-guest adsorption interactions with the elastic response of the three-dimensional framework of a given geometry. The first results obtained using this model were showcased in an earlier letter,[11] and we aim here at giving its full description as well as more recent results we obtained from it. The structure of the paper is as follows. In section II.A, we analyze the basic equations of the elasticity theory in three-dimensional frameworks. Special attention is paid to the compatibility conditions implied by the Saint-Venant principle. We conclude in section II.B that the deformation in MIL-53 type crystals occurs as concerted shear motion of 2D layers of cells. As such, the transition can be described with one principal order parameter related to the cell volume. This reduces the initial three-dimensional model to a one-dimensional model with long-range elastic interactions. In section II.C, we show how to build a minimalistic model Hamiltonian coupling local elastic deformations, long-range elastic interactions and guest adsorption. The latter is described by the adsorption energy and adsorption stress. This system dynamics is analyzed by performing Grand Canonical Monte Carlo simulations (sections II.D and II.E) modified to account for the energy barrier of the lp$\leftrightarrow$np transition. The specifics of the transition dynamics are studied in section III by varying the energy barrier and the parameter responsible for elastic interactions.

We also explore the influence of the crystal size on the transition dynamics. In section IV, summarizing the results, we conclude that the suggested model explains the mechanisms of breathing transition and reproduces its experimental features including structural changes, hysteretic nature of transitions, and phase coexistence.

## II. Model description

### *II. A. General treatment of deformation and elastic compatibility equations*

In order to introduce the elastic forces within the three dimensional lattice of a certain crystallographic symmetry, we will follow below a rigorous approach of the theory of elasticity based on the deformation strain tensor $\varepsilon$ defined on the level of individual unit cells. This tensor characterizes the states associated to lp and np phases as well as the intermediate structural configurations along the path of the transition phase between them. In general, the deformation strain tensor $\varepsilon$ between the two states 1 and 2 is defined through the difference of the metric tensors, $g^1$ and $g^2$, associated with these states:

$$\varepsilon = \frac{1}{2}\left(g^2 - g^1\right) \qquad (1)$$

The metric tensor is a rank 2 tensor which can be calculated from scalar products of the generating Bravais lattice vectors $\{ \vec{v}_i, i = 1, 2, 3 \}$ as such:

$$g_{ij} = \vec{v}_i \cdot \vec{v}_j \qquad (2)$$

The components of the deformation strain tensor are not independent. In three dimensions, the displacement field has 3 independent components at any point, and the symmetric strain tensor nominally has *6* independent components. Because the strain tensor is composed of the derivatives of the local displacement vector field, there exist inherent relationships, or constraints, among its components. These constraints are imposed by the compatibility equations,[12,13,14,15,16]

$$\sum_{\Gamma} \varepsilon = 0 \qquad (3)$$

Being $\varepsilon$ the deformation strain, and $\Gamma$ any closed path over the crystal lattice lattice. These equations preserve the cell lattice integrity and reduce the number of parameters needed for the quantitative description.[17,18,19,20,21] The non-locality of these constrains gives rise to the long-range elastic interactions in the lattice. Thus, even in the case of near-neighbor interactions, eq. (3) implies non-local interactions on the level of the lattice.

### *II.B. Framework geometry and elastic compatibility conditions*

Parametrization of the deformation strain tensor involves the selection of primary and secondary order parameters (OP). The primary OP quantifies the path along the phase transition, while secondary OPs describes other modes of the elastic response of the material. We will show later that these OPs are the suitable variables for the description of the framework's free energy. This free energy is a function of the scalar products of the lattice basis vectors, or elements of the lattice metric tensor. In general, the free energy is expressed through a certain polynomial of these variables[22,23,19,13], that can be further approximated by (i) keeping the terms involving only the primary OP; (ii) approximating the dependence on secondary OPs by quadratic terms, complying with the standard linear elastic behavior; and (iii) neglecting all other terms. In this section, we adopt this approach to the particular problem considered here.

From the geometrical standpoint, the lp and np phases of MIL-53 differ mainly by the shape of cells, which can be characterized by the angle $\theta$ of its rhombus cross-section; in lp phase, $\theta \sim 79°$, while in np phase, $\theta \sim 40°$ (Fig. 2). That means that the primary OP can be described by a scalar related to this angle. At the same time, the variation in the linker length $a$ between the two phases is of the order of 0.4 Å (3.8%), making the length of the rhombus sides *essentially* invariant upon deformation of the structure. In the model presented here, we will make the assumption that the linker length, $a$, as well as the unit cell length in the orthogonal direction, $b$, are constant and thus we do not consider any secondary OP. This assumption implies that the deformation of a single cell can be quantified by just one degree of freedom represented by the rhombus angle $\theta$, which imposes a strong constraint on the framework geometry. This conclusion can be formally derived from elastic compatibility conditions, eq. (3), linking the deformation strain in neighboring cells[24]. Indeed, the normalized lattice vectors are $\vec{v}_1 = \vec{e}_x$, $\vec{v}_2 = \cos\theta\,\vec{e}_x + \sin\theta\,\vec{e}_y$, and $\vec{v}_3 = \vec{e}_z$. Thus, the only variable components of the metric tensor are $g^2_{12} = \cos\theta$ and, $g^1_{12} = \cos\theta_{\mathrm{lp}}$. Thus, according to eq. (1), the only non-zero strain tensor component is the shear component $\varepsilon_{12}$,

$$\varepsilon_{12} = \frac{1}{2}\left(\cos\theta - \cos\theta_{\mathrm{lp}}\right), \qquad (4)$$

which represents the primary and unique scalar OP.[25] The undistorted phase in this representation corresponds to the lp phase characterized by $\varepsilon_{12} = 0$.

It is convenient to define a symmetrized strain $e$, as

$$e_i = -1 + 2\frac{\cos\theta_i - \cos\theta_{\mathrm{lp}}}{\cos\theta_{\mathrm{np}} - \cos\theta_{\mathrm{lp}}} \qquad (5)$$

in order to have a symmetric and normalized representation for the equilibrium values of the strain, that is $e = -1$, for the reference lp phase, and $e = +1$ for the reference np phase. The cell volume can be expressed in terms of the symmetrized deformation strain as $V_i = a^2 b\sqrt{1-\left(\frac{\cos\theta_{\mathrm{np}}-\cos\theta_{\mathrm{lp}}}{2}+\frac{\cos\theta_{\mathrm{np}}-\cos\theta_{\mathrm{lp}}}{2}e_i\right)^2}$, where $a$ is the linker length and $b$ is the length of the unit cell vector along the channel (orthogonal to the channel cross-section).

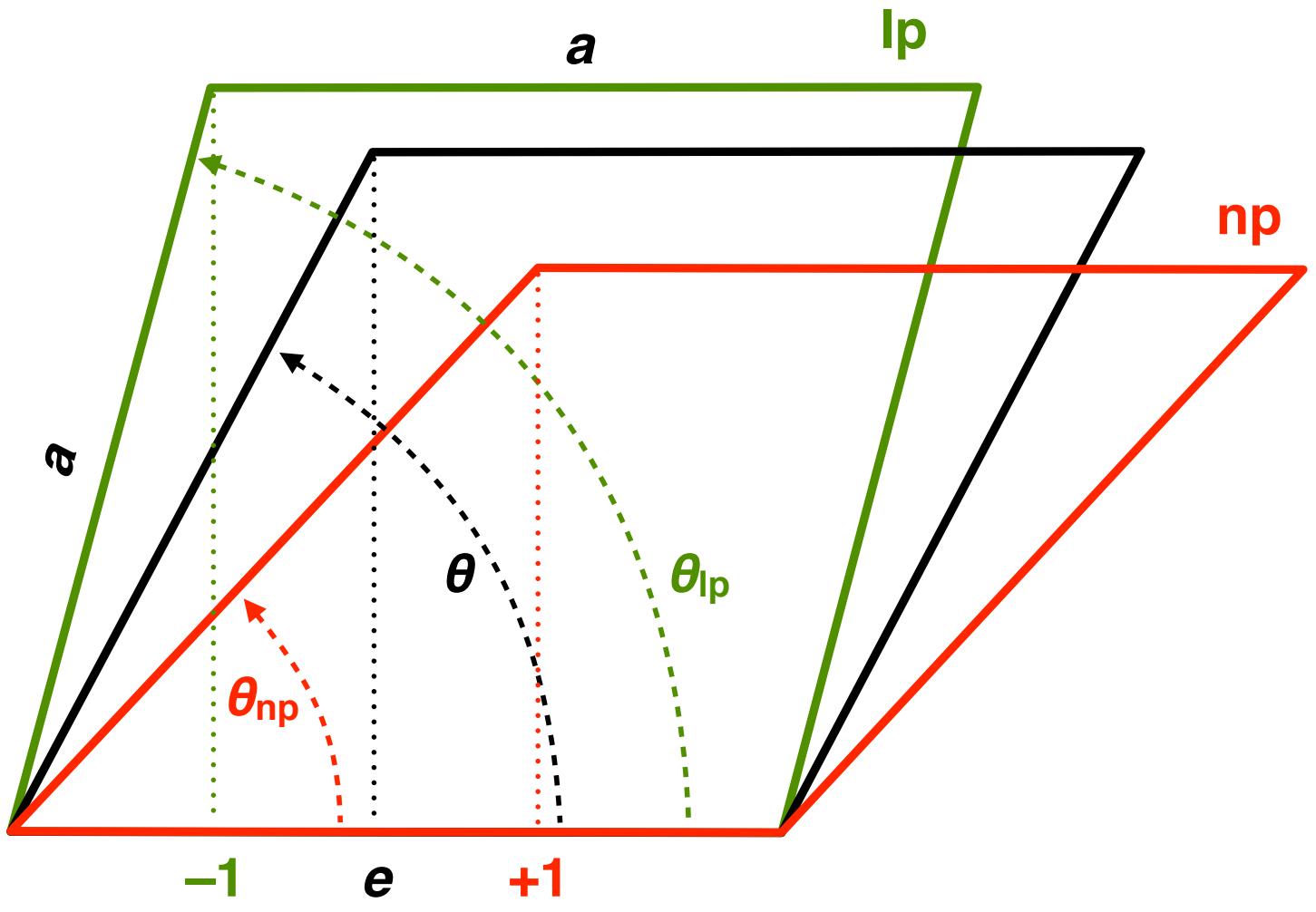

**Fig. 2** Schematic diagram of the cell cross-section for the structural phase characterization. The green and red frames correspond to the lp ($e = -1$) and np ($e = +1$) phases, respectively. The linker length $a$ is constant.

Elastic compatibility equations imply that the sum of deformations along any close trajectory (loop) in the cell network should be null (Eq. 3). As such, the compatibility equations for three independent loops Γ in the XZ, and ZY planes of the cell lattice (depicted on Fig. 3) can be written as,

$$\sum_{\Gamma_{XZ}} e_{i,j,k} = 0 : a + \sqrt{a^2 - e_{i+1,j,k}^2} - a - \sqrt{a^2 - e_{i,j,k}^2} = 0 \Rightarrow e_{i,j,k} = e_{i+1,j,k}$$
$$\sum_{\Gamma_{ZY}} e_{i,j,k} = 0 : a + e_{i,j+1,k} - a - e_{i,j,k} = 0 \Rightarrow e_{i,j,k} = e_{i,j+1,k} \quad , \quad (6)$$

Trivial equations for the YX plane are omitted because they do not bring any additional constraint. Eqs. 6 express the constraint on the derivatives of the deformation strain field. Gradients of the deformation strain field must be zero at any point over the lattice in the directions *x*, and *y*, this can be expressed as,

$$D_x e_{i,j,k} = 0$$
$$D_y e_{i,j,k} = 0 \quad (7)$$

Here $D_i$ represents the standard discrete partial derivative operator, e.g. $D_x e_{i,j,k} = e_{i+1,j,k} - e_{i,j,k}$. Eqs. (7) can be easily translated in terms of the cross-section angle $\theta$, as $D_x \theta_{i,j,k} = 0$, and $D_y \theta_{i,j,k} = 0$. This condition requires the constancy of $\theta$ within the XY-plane that reduces the model to one dimension, since the chosen OP can only vary in Z-direction.

As such, the local elastic compatibility equations give rise the long-range elastic interactions, which impose a firm constraint on cell deformations requiring the similitude of the cell shapes within the 2D layer of cells in the XY-plane. This means that the phase transformations occur in a cooperative manner and necessarily involve the entire layer of cells, all of which must be in the same phase. As a consequence, the layer of cells, rather than the single cell, is to be taken as a basic unit for the framework mechanical model. As such, the 3D framework can be presented as a 1D stack of 2D layers of identical cells. This makes possible to formulate a 1D minimalistic model of the framework deformation that captures the main system properties with a minimum number of input parameters.

The above conclusion is the key for the simulation analysis presented in this paper, and it is worth of additional discussion in terms of the variation of cell geometry during the phase transformation. The assumption of the preserved rhombus shape with the fixed side length imposes a strong correlation on the deformation of neighboring cells in order to ensure the lattice integrity[14]. First, the cells along a channel must coherently deform, as two stacked cells of rhombus shape cannot match unless they have the same angle $\theta$. Secondly, the channels connected by sides of fixed length must possess equal $\theta$ along the shear direction. This defines a 2D layer of cells, inside which all the channels have the same cross section. Mismatch in the rhombus angle can exist along the direction perpendicular to the shear plane. Thus, the only possible mechanism of framework deformation is the layer-by-layer shear and the elementary deformation consists of the shear of the layer of cells in the direction perpendicular to the channel axis represented in Fig. 3.

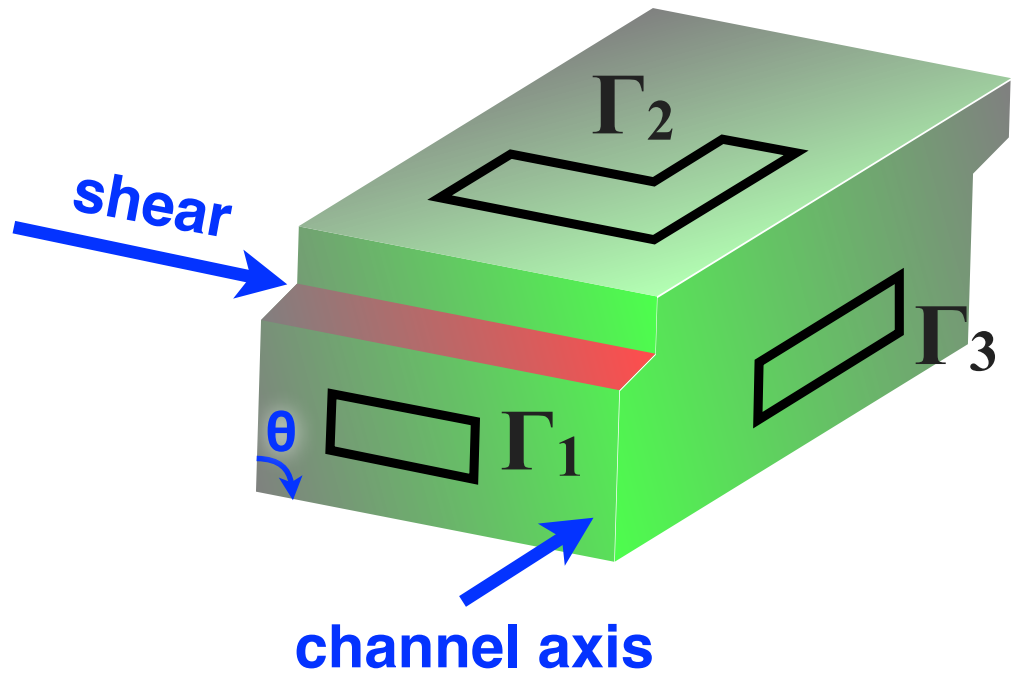

**Fig. 3** Schematic representation of the MIL-53 framework after the first event of lp-np transformation, which involves in-plane shear of a 2D layer of cells in a direction orthogonal to the channel axis.

### *II. C.* 1D *model of adsorption in bi-stable framework*

Within the proposed 1D model, the two main variables describe the state of each cell layer at given external thermodynamic conditions. First is the deformation state of the material, given by the stain field $e$. The second variable is the adsorption loading $n$, or the mean number of guest molecules per unit cell in the layer. By averaging the loading over the whole framework as a function of the external gas pressure (or chemical potential), one obtains the adsorption isotherm. This quantity is the key observable of the system, as it is measured in isothermal adsorption experiments. The experimental adsorption isotherms typically display two sharp yet continuous transitions from almost empty lp phase to almost fully loaded np phase and from np phase to loaded lp phase, as shown in Fig. 1. The loading capacities of both phases differ significantly and represent the main quantitative parameters determining the adsorption isotherm behavior. In the proposed minimalistic model, we assume the loading is discrete and allow for adsorption of either, 0, 1, or 2 molecules in the cell, that correspond to empty cell, the np cell capacity, and the lp cell capacity. This simplification can be easily generalized by introducing additional loading levels, or by considering adsorption loading as a continuous variable.

The proposed model describes the interplay between guest adsorption and host framework deformation in terms of a Hamiltonian that depends on the loading ***n*** and strain ***e*** fields and is expressed per unit cell. The Hamiltonian contains two main terms. The first term accounts for the host energy, while the second expresses the interaction energy between the host and the guest particles,

$$H(n,e) = H_{\mathrm{host}}(e) + H_{\mathrm{host-guest}}(n,e) \qquad (8)$$

The explicit form of the host energy is:

$$H_{\mathrm{host}}(e) = \sum_i \left[ \frac{c_0}{2}\left(e_i - s_i\right)^2 + \frac{\Delta F}{2} s_i + \frac{c_i}{2}\left(De_i\right)^2 \right] \qquad (9)$$

The first term in the right-hand side accounts for elastic deformation of individual cells where the local elastic energy is modeled as by two parabolic potentials around the undistorted np and lp states,[26] see Fig. 4; variable $s_i$ is discrete and denotes the phase state of the cell ($s_i = -1$ in lp phase and $s_i = +1$ in np phase), and $(e_i - s_i)$ is the local deformation from the reference undistorted state of the respective phase. The parabolic elastic potential wells are effectively cut to avoid the unrealistically high energies of intermediate states by introducing the free energy barrier $E_B$ that should be crossed in the course of the phase transformation as discussed in more details in Section II.E. The effective elastic constant $c_0$ is assumed equal for both phases just for the sake of simplicity. The differences in the elastic constants of np and lp phases[27] can be introduced in a more complex version of the model. The second term $\Delta F$ represents the difference in the free energy between empty np and empty lp phases, which is positive, reflecting the fact that the initial "dry" state corresponds to the stable lp phase. This energy was estimated within the thermodynamic model developed in Ref. 7. The third term corresponds to the non-local cell–cell elastic energy proportional to the strain gradient squared; it penalizes the formation of interfaces between layers of lp and np phases ($c_1 > 0$), by the interfacial energy of $2c_1$, and also levels elastic deformations in neighboring cells of the same phase.

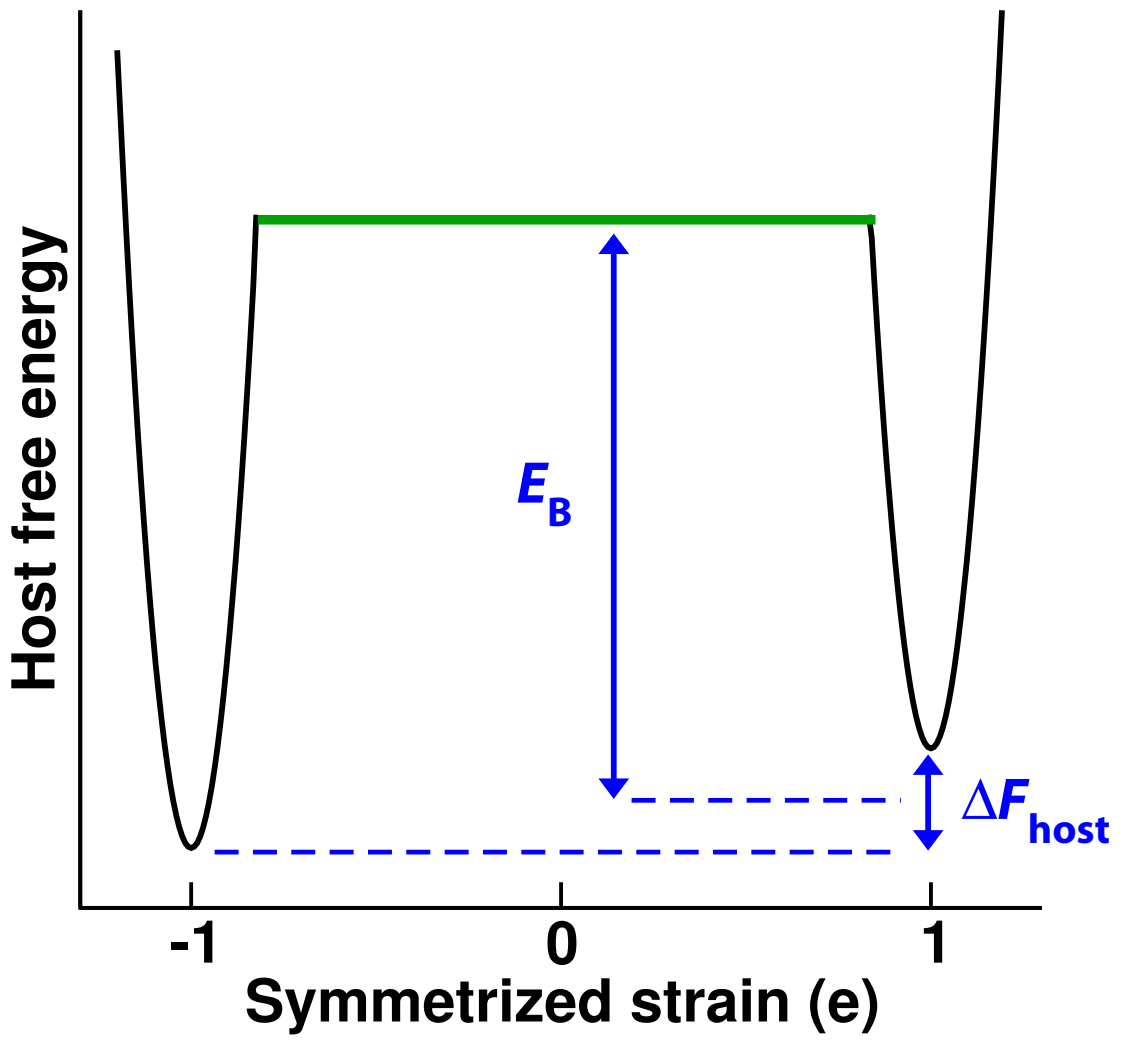

**Fig. 4** Schematics of the host free energy landscape for the "dry" bi-stable framework, as a function of symmetrized strain, *e*. Two regions of elastic deformation around the equilibrium lp and np structures ($e = \pm 1$) are approximated by parabolas. The non-elastic region in-between is taken into account by introducing a free energy barrier $E_B$ in the dynamic model.

The *host-guest* energy is expressed as a sum of two terms,

$$H_{\text{host-guest}}(e) = \varepsilon_a(n_i, s_i) - \sigma_a(n_i, s_i)\left(e_i - s_i\right) \quad (10)$$

which determine the host-guest interactions as the energy of adsorption $\varepsilon_a(n_i, s_i)$ at given loading $n_i$ and deformation $e_i$ expanded around the adsorption energy $\varepsilon_a(n_i, e_i)$ in non-deformed reference lp or np state, $e_i = s_i$. This expression gives rise to a quantity of prominent physical significance: the adsorption stress induced on the host framework due to its interactions with the guest molecules, which is defined as

$$\sigma_a(n_i, s_i) = \left.\frac{\partial \varepsilon_a(n_i, e_i)}{\partial e_i}\right|_{e_i = s_i, n_i} \quad (11)$$

in line with the thermodynamic definition of the adsorption stress. The adsorption stress couples the host-guest interactions with the elastic deformation and accounts for the forces exerted by the guest molecules on the host framework. It can be either negative or positive depending on the loading and thus cause either elastic contraction or expansion. As such, the number of input parameters characterizing adsorption in our model is reduced to 8: 4 adsorption energies $\varepsilon_a(n_i, s_i)$ and 4 adsorption stresses $\sigma_a(n_i, s_i)$, $n_i = 1$ or 2, $s_i = \pm 1$.

### *II. D. Discretization of the model.*

Modeling the dynamics of the coupled adsorption and deformation in the process of incremental stepwise variation of the chemical potential of the adsorbate, we make a further assumption that the local elastic relaxation of the framework occurs on a much smaller time scale than the establishment of adsorption equilibrium. Under this assumption of fast relaxation of the elas-tic degrees of freedom, the total energy should be at its minimum with respect to the stain field ***e***. As such, the quasi-equilibrium strain field can be found from the minimization of the Hamiltonian at given discrete state and loading fields, ***s*** and ***n*,**

$$\left.\frac{\partial H(\{n_i\},\{e_i\})}{\partial e_k}\right|_{\{n_i\},\{s_i\}} = 0 \quad \forall k \qquad (12)$$

Since the second derivative is always positive and equal to $c_0 > 0$,

$$\left.\frac{\partial^2 H(\{n_i\},\{e_i\})}{\partial e_k^2}\right|_{\{n_i\},\{s_i\}} = c_0 \quad \forall k \qquad (13)$$

the condition of the minimum is trivially satisfied, and thus defined strain field would correspond to a mechanically stable state.

Eq. (12) implies zero total stress within the framework while the adsorption stress, eq. (11), can be different from zero. The frame must deform in order to satisfy this condition. The assumption of the local elastic equilibrium allows us to determine the continuous elastic strain field ***e*** as a function of the discrete fields, ***s*** and ***n*.** That is, the dynamics of the system is governed by the evolution of the discrete field ***s*** and ***n*** that allows for the further simplification of the model. Due to the quadratic nature of the elastic potential, the minimization, eq. (12), yields a system of linear equations, which can be solved in a matrix form to determine the strain field at given distribution of phases $\{s_i\}$ and loadings $\{n_i\}$,

$$e_i = \sum_{k=1}^{L} M_{ik}\left(c_0 s_k - \sigma_a(n_k, s_k)\right), \text{ where } M_{ik} \equiv (c_0 I_{ik} + c_1 H_{ik})^{-1}. \qquad (14)$$

Here, $I_{ik}$ is the identity matrix, and $H_{ik}$ is a suitable matrix representation of the strain gradient term in eq. (9), i.e. $H_{ik} \equiv I_{i,k} - I_{i+1,k} - I_{i,k+1}$.

Eq. (14) allows one to express the Hamiltonian given in eq. (9) and eq. (10) in terms of the discrete fields ***n***, and ***s***,

$$H^*(\mathbf{n},\mathbf{s}) = \frac{1}{2}\sum_{i,j=1}^{L}\left[c_0^2 J_{ij} s_i s_j - M_{ij}\sigma_a(n_i,s_i)\sigma_a(n_j,s_j) - 2c_0 J_{ij} s_i \sigma_a(n_j,s_j)\right] + \sum_{i=1}^{L}\left[\frac{\Delta F}{2} s_i + \varepsilon_a(n_i,s_i)\right] + \frac{c_0 L}{2} \quad (15)$$

Here, $J_{ij} = I_{ij} / c_0 - M_{ij}$. As such, within the discretized model, eq. (15), each cell layer is characterized by its state variables, $s_i = \pm 1$, and loading $n_i$ = 0, 1 or 2. Note that eq. (15) explicitly accounts for the long-range coupling between adsorption states of different loading in $M_{ij}\, \sigma_{\mathrm{a}}(n_i,s_i)\, \sigma_{\mathrm{a}}(n_j,s_j)$ term; such coupling term was not present directly in the initial Hamiltonian, eq. (10), but it arises from the elastic interactions due to adsorption-induced stress and compatibility conditions.

### *II. E. Dynamics of the system*

In modeling the adsorption process, we performed grand canonical Monte Carlo (MC) simulations of the 1D array of cell layers with open (non-periodic) boundary conditions. The external variables are the size of the system, *L*, the temperature *T,* and the chemical potential $\mu$ of the adsorbate. Two types of MC moves, which lead to the change of the cell state, are considered: adsorption or desorption of one particle ($n_i \rightarrow n_i \pm 1$) and the phase switch between lp and np structures ($s_i \rightarrow -s_i$). The acceptance probabilities *W* for particle insertions and deletions are standard for the Metropolis MC scheme:

$$W(n_i \rightarrow n_i \pm 1) = \min\left\{1, \exp\left[-\beta(\Delta H^* \mp \mu\right]\right\} \quad (16)$$

Where $\beta = 1/k_B T$, and the energy difference $\Delta H^*$ has the following explicit form:

$$\Delta H^* = -\Delta\sigma_i \left( \sum_{i,j=1;i\neq j}^{L} M_{ij}(\sigma_j^a + c_0 s_j) + c_0 J_{ii} s_i \right) - \frac{1}{2} M_{ii} \left(\Delta\sigma_i^a\right)^2 + \Delta\varepsilon_i \tag{17}$$

The phase switch attempt implies the transition from one elastic potential well to the other across the energy barrier $E_B$, see Fig. 3. The probability of barrier crossing is described by the following the scheme employed by Kang et al.[28] in studies of activated diffusion. As such, we introduce the energy barrier directly in the acceptance probabilities of the phase change.

$$W(s_i \to -s_i) = \begin{cases} \min\left(1, \exp\left[-\beta\Delta H^*\right]\right) & \text{if } \left|\Delta H^*/2\right| > E_B \\ \exp\left[-\beta\left(\Delta H^* + \dfrac{E_B}{2}\right)\right] & \text{if } \left|\Delta H^*/2\right| < E_B \end{cases} \tag{18}$$

The energy difference associated to the phase change move is given by,

$$\Delta H^* = -c_0 \Delta s_i \left( \sum_{i,j=1;i\neq j}^{L} M_{ij}(\sigma_j^a - c_0 s_j) + J_{ii}\sigma_i \right) + \Delta\varepsilon_i - s_i \Delta F \tag{19}$$

The probability of phase transition is the same as in the standard MC scheme presented in eq. (16) provided $|\Delta H^*/2| > E_B$. Otherwise, eq. (18) introduces a possibility for metastable states in the system and, as a consequence, for hysteretic behavior. This statement is visualized in Fig. 5, where we schematically show all the different possibilities of the transition probability.

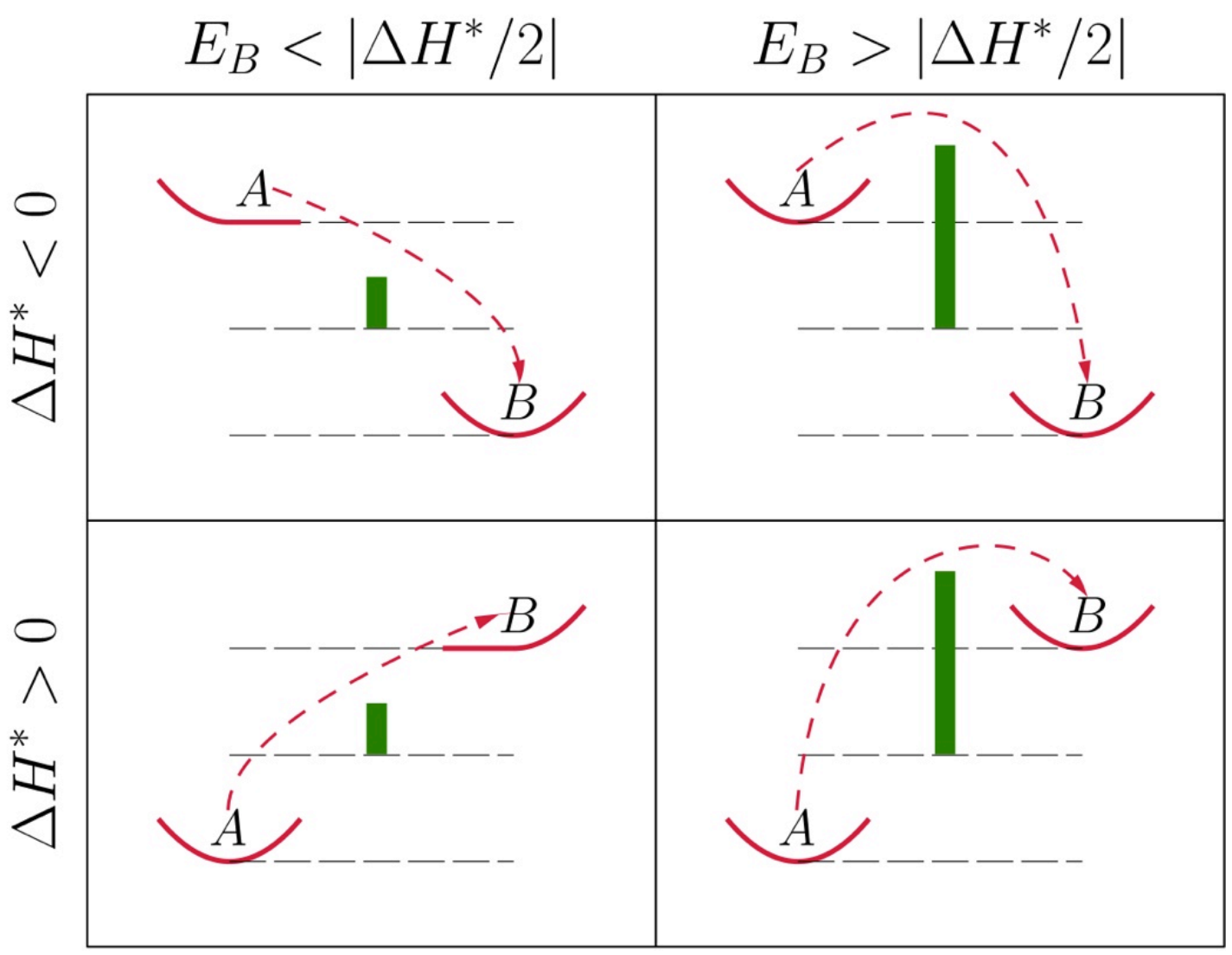

**Fig. 5** Schematics representation of the transition between the states and the effect of the energy barrier. When the barrier is low, the transition probability is determined from the standard Metropolis scheme (left panels). When the barrier is high, the transition probability is modified according to eq. (18) (right panels).

Left panels in Fig. 5 correspond to low barriers that do not affect the transition. Right panels illustrates the possibility of the metastable states separated by the barrier provided the latter is high enough. In this case, even when the transition is associated with the energy reduction, the move is accepted not with a probability equal to one, as in the Metropolis scheme, but with the probability that accounts for the barrier crossing.

## *III. Results and discussion*

### *III.A. Obtaining main features of breathing transitions.*

The minimalistic model reproduces on a qualitative level the main features of breathing transitions presented above in Fig.1, namely the double transition upon adsorption, first from lp to np and then from np to lp phases. In the examples of calculations presented below, the adsorption energies and stresses are chosen to qualitatively reproduce adsorption isotherms in each phase, taking as a reference 200 K Xe adsorption data.[7] As such, each individual adsorption isotherm is of the IUPAC type I and shows a Langmuir-like behavior. The transitions imply gradual transformation of the sample from one phase to the other, mechanisms of which are discussed below. As shown in Ref. 29, the stress developed in individual phases upon adsorption and desorption is non-monotonic, featuring contraction and expansion in the course of adsorption.

Choosing the model parameters, we grouped them in two categories. The first group of parameters is based on the experimentally measurable properties, such as elastic constant $c_0$ (i.e. bulk modulus), adsorption energies (adjusted to reproduce semi-quantitatively the experimental isotherms), adsorption stress (their sign is known but not their value; however, their exact values are not crucial to the conclusions of the model). These parameters are necessarily correlated in order to reproduce the phenomenology (adsorption energy larger in np phase that in lp phase, etc.). These parameters are presented in Table 1, and they are fixed in all following calculations.

| | | | |
|---|---|---|---|
| $c_0 = 100$ | | $\Delta F = 5.0$ | |
| $\varepsilon(1,\mathbf{lp}) = 0$ | $\varepsilon(1,\mathbf{np}) = -10$ | $\varepsilon(2,\mathbf{lp}) = -2$ | $\varepsilon(1,\mathbf{np}) = 3$ |
| $\sigma_a(1,\mathbf{lp}) = -10$ | $\sigma_a(1,\mathbf{np}) = -10$ | $\sigma_a(2,\mathbf{lp}) = 10$ | $\sigma_a(1,\mathbf{np}) = 10$ |

**Table 1** Values of the fixed parameters in units of $k_B T$.

The parameters given in Table 1 ensure that at the condition of thermodynamic equilibrium, the adsorption process will possess double breathing transition. The equilibrium isotherm can be calculated on the level of one cell layer ignoring the dynamic effects related to the inter-layer coupling and to the energy barrier of phase transitions. The model parameters responsible for the dynamics and hysteretic behavior represent the second group. They include the layer-layer elastic coupling $c_1$, free energy barrier $E_B$, as well as boundary conditions and system size. These parameters are not directly related to the experimental observations and are varied in section II.B and II.C to demonstrate their impact on the mechanism of breathing transitions.

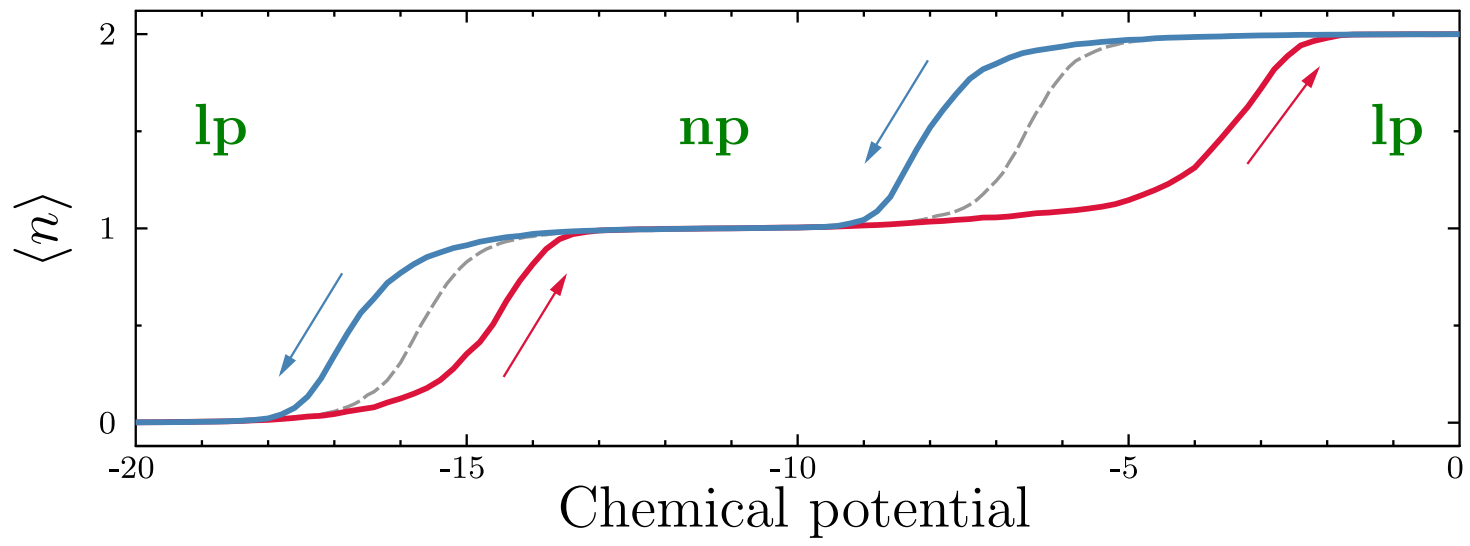

**Fig. 6:** Principal features of breathing transitions: adsorption isotherm (red line), desorption isotherm (blue line), reversible isotherm obtained by ignoring the energy barrier $E_B$ in simulation (gray). Model parameters from Table 1, coupling parameter $c_1 = 1$.

In Fig. 6, we present typical results of simulation for $L = 2000$, $E_B = 10.5$ , and $c_1 = 1$ in comparison with the equilibrium isotherm determined at $E_B = 0$ , and $c_1 = 0$. The adsorption isotherms are given as the average loading $\langle n \rangle$ *vs* the chemical potential $\mu$. The adsorption and desorption isotherms (top panel) display a prominent hysteresis similar to the experimentally observed, see Fig.1. The hysteresis is also apparent for the stress isotherms $\langle \sigma_a \rangle$ (not shown here), and the evolution of the sample composition characterized by the fraction of layers in lp phase $x_{lp}$.[11]

### *III.B. Effect of elastic coupling and free energy barrier of phase transition*

We first consider the system of large size, i.e. in the thermodynamic limit. In practice, this is achieved at the length $L = 1000$, which was determined by comparison of the results for systems of increasing size.

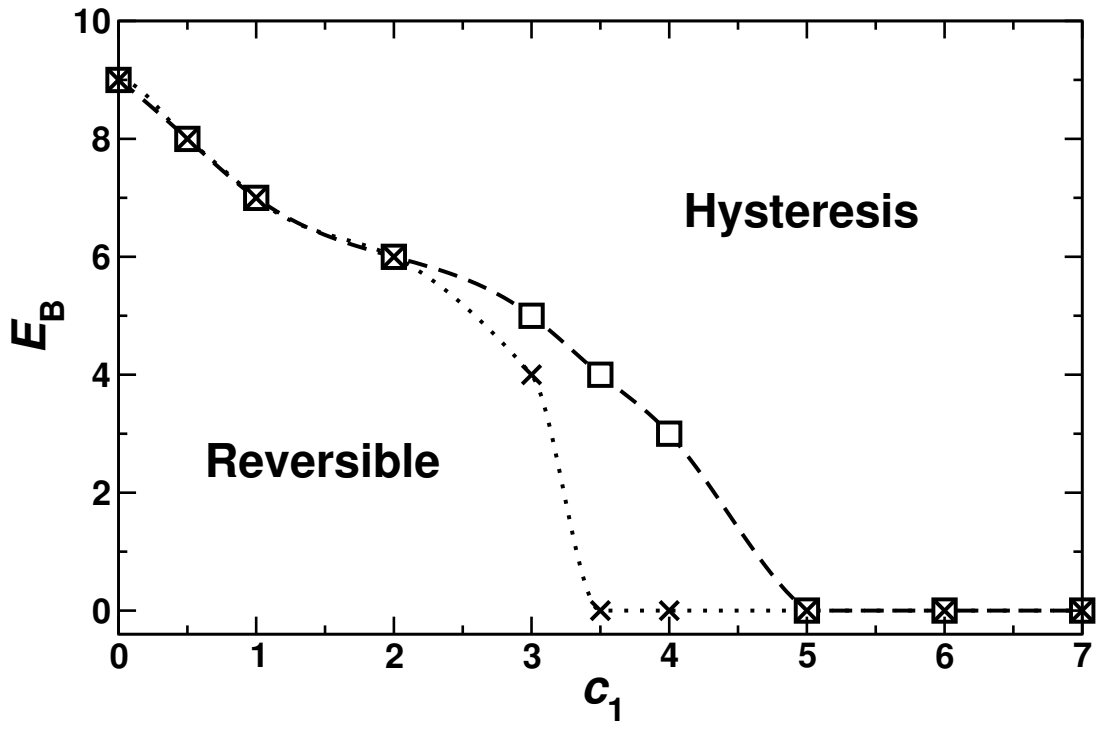

**Fig. 7:** Presence of hysteresis in breathing transitions depending on model parameters $E_B$ and $c_1$, at the thermodynamic limit. Dotted and dashed lines represent the limits of the reversible and hysteretic regimes for the low-pressure and high-pressure transitions, respectively.

The hysteretic behavior is not observed for all values of the layer–layer elastic interaction parameter $c_1$ and the energy barrier $E_B$. We performed a systematic study of the region of hysteresis by varying parameters $E_B$ and $c_1$ and constructing a "phase diagram" separating the equilibrium and hysteretic regions in parameter space, Fig. 7. In the absence of both barrier and elastic coupling, the breathing is fully reversible. By introducing either a large enough layer-layer coupling, or a large enough energy barrier, or both, the structural transitions become hysteretic. It can be seen that the boundaries of nonreversible behavior for the two structural transitions (low-pressure, represented as a dotted line, and high-pressure, represented as a dashed line) are quite close to each other. As such, it is possible yet unlikely that in the case of two breathing transitions, the low-pressure transition would be reversible and the high-pressure one hysteretic.

The main conclusion we can draw from Fig. 7 is that both the free energy barrier and the layer-layer elastic coupling can cause hysteretic structural transformations, which are observed in all known experimental occurrences of breathing transitions. While it is clear the energy barrier stabilizes the presence of metastable states, the effect of the elastic coupling is less straightfor-ward. However, the interlayer coupling (stemming from cell-cell elastic interactions) penalizes the switch of a single layer of cells in the material, and thus plays in the system dynamics a similar role to the free energy barrier.

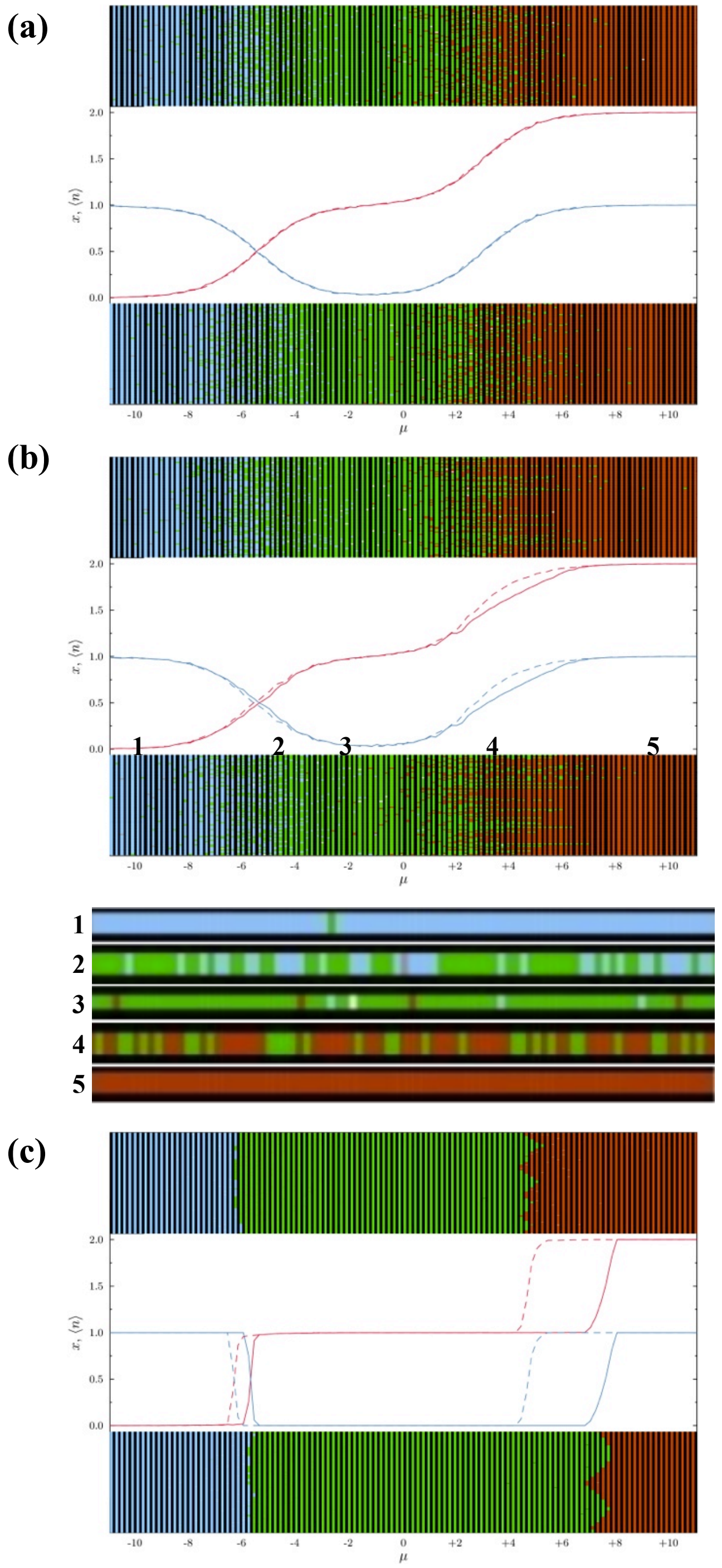

**Fig. 8.** Adsorption isotherms, $\langle n \rangle$ (red curves); phase fraction, $x_{\mathrm{lp}}$ (blue curves) as a function of chemical potential $\mu$. On top and bottom are represented spatial distribution of phases and loadings, for adsorption (bottom) and desorption (top), at corresponding $\mu$. Each vertical bar represents the 1D succession of 2,000 layers colored with respect to their states: blue - empty lp phase; grey - empty np phase, green - np phase with one particle; white - lp phase with one particle; red - lp phase with two particles; black - np phase with two particle (not observed in these simulations). a : $c_1 = 0$, $E_{\mathrm{B}} = 0$; b: $c_1 = 0$, $E_{\mathrm{B}} = 9$; c: $c_1 = 8$, $E_{\mathrm{B}} = 0$. Panel b includes a zoom on characteristic phase distributions in the system for five labeled points along the adsorption isotherm.

In order to characterize the nature of the mechanism of the hysteretic transition, we present in Fig. 8 the adsorption and desorption behavior for three characteristic sets of parameters: top panel (a) no layer-layer coupling and no energy barrier, (b) with only layer-layer coupling, no energy barrier, and (c) with only a free energy barrier, no coupling. In each case, a graphical representation of the system upon adsorption and desorption is plotted below and above the isotherm. Our "reference" case here will be the reversible simulation with $c_1 = 0$ and $E_{\mathrm{B}} = 0$. In this case, the reversible isotherm stems from a smooth transition of layers from one state to another; the phase distribution in the regions corresponding to the structural transitions appear as random intermittent domains of lp and np layers. From this no-coupling no-barrier case, the introduction of a high enough barrier ($E_{\mathrm{B}} \geq 9$) slows down the dynamics of the system and allows the creation of "metastable" states, in which domains of the new phase nucleate and grow (panel b of Fig. 8). This is reflected in the hysteretic nature of the transition. Below this panel we show zooms on 5 characteristic distributions of phases, obtained at labeled point 1–5 along the adsorption isotherm: the first one shows an empty system in lp phase, with a fluctuation of a single cluster of np phases; the second one corresponds to lp-np coexistence; the third one is located between the two adsorption steps, where the np phase predominates but some inclusions of empty lp phase and full lp phase are seen; the fourth case is np-lp coexistence in the region of the second transition, and the last distribution shows the final state of adsorption, a system of fully loaded lp lay-ers. In the presence of large elastic layer-layer correlation, and to some extent in the case of high energy barriers, hysteresis loops widen as the nucleation rate for the phase transition is reduced. For example, in the case of $E_B = 0$ and $c_1 = 8$ (lower panel of Fig. 8), we see that only a few nucleation events occur in the entire system (3 events upon adsorption, 4 upon desorption), yielding fairly steep steps in both the loading and phase composition isotherms (avalanche effect).

**III.C. Effect of crystalline domain size on breathing dynamics**

The examples discussed above were calculated for a large system comprised of $L = 2000$ layers. For such large system, no difference in adsorption hysteretic behavior was observed with periodic and free boundary conditions. Except for the very last example of the strongest elastic coupling, Fig. 8c, the distribution of the new phase nucleation events was uniform along the system with free boundary conditions. However, for the adsorption np-lp transition, the influence of the boundaries is apparent: one may see only three nucleation events, two at the boundaries and one in the center, and an avalanche phase growth between them. As the system size decreases, the effect of system boundaries becomes stronger. To study the influence of the system size on the transformation dynamics, we performed simulations of the systems of various sizes $L$ with free boundary conditions.

In Fig. 9, we demonstrate the adsorption isotherms in the systems of different sizes for two typical cases of strong ($c_1 = 6$, left panel) and weak ($c_1 = 2$, right panel) interlayer elastic coupling. The energy barrier, $E_B = 9$, is chosen high enough to secure the hysteretic behavior in the thermodynamic limit even in the absence of elastic interaction penalty. For small size system, the boundary effect is important. Indeed, the nucleation of the new phase at the boundary implies a smaller overall free energy penalty, which is a cumulative effect of the energy barrier and inter-layer coupling. The latter effect is approximately twice as small for the boundary layer than for the internal one. As such, the phase transformation tends to start from the boundary and propagates to the center of the system, as shown in Fig. 10 for np-lp transition in the strong coupling case. With the decrease of the system size, the hysteresis progressively shrinks. For the low coupling case, this effect is more pronounced. The hysteretic behavior is observed in the thermodynamic limit only. For all system sizes including and below 100, the isotherm is reversible.

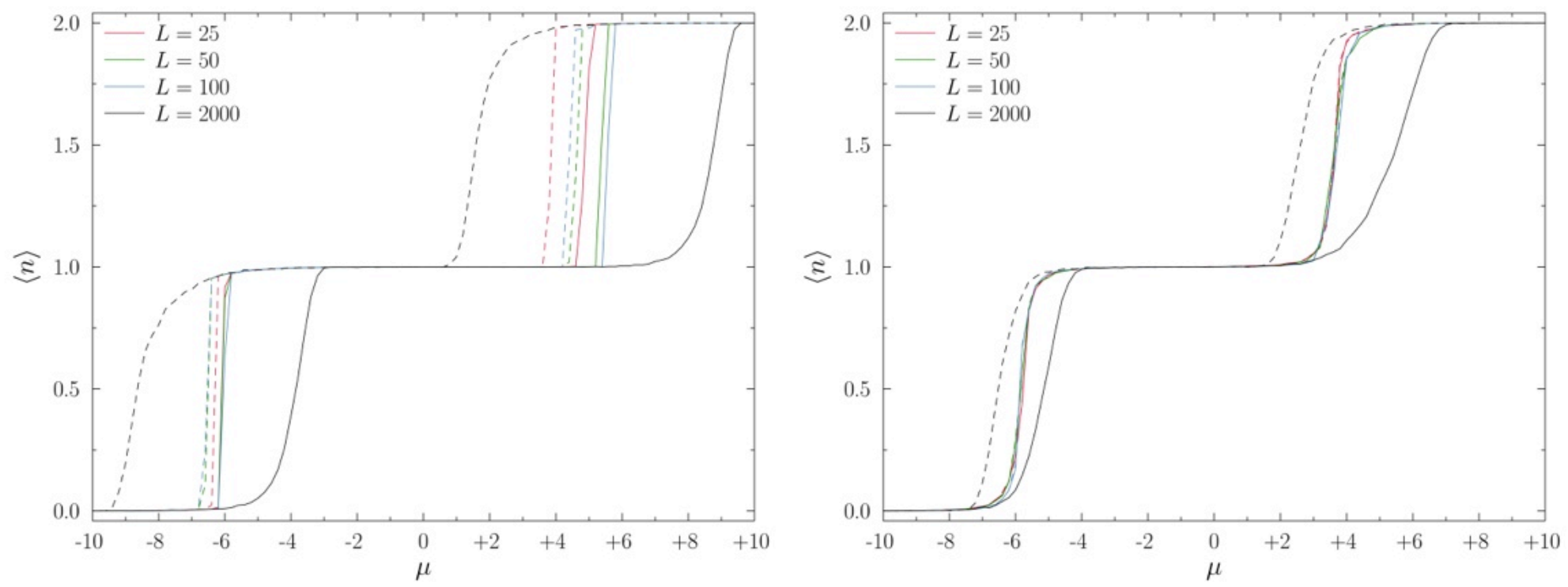

**Fig. 9:** Influence of system size ($L$ = 25, 50, 100 and 2000) on the dynamics of breathing transitions for two dynamic regimes. Left - strong elastic coupling; right - weak elastic coupling. Hysteresis increases with the system size.

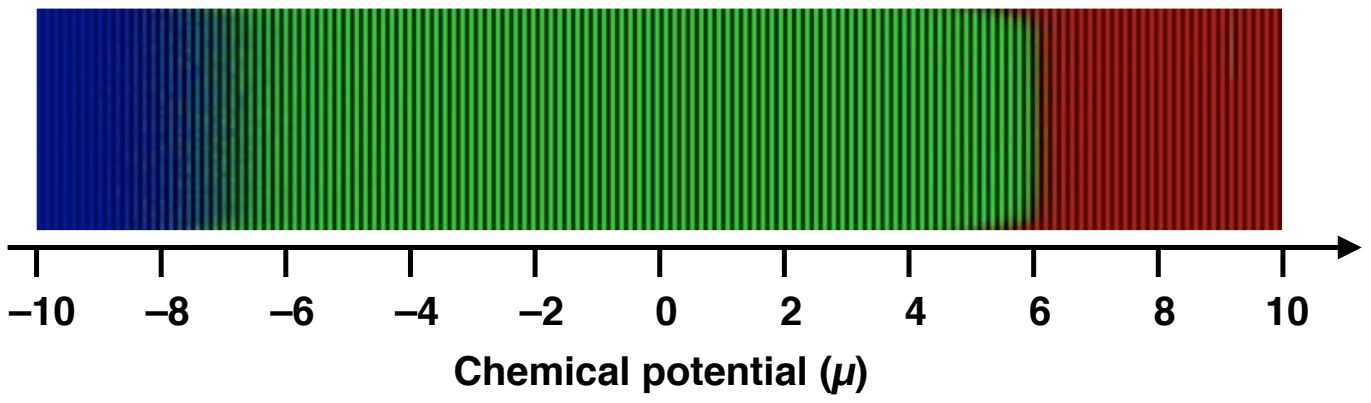

**Fig. 10** Phase distribution upon adsorption at $L = 100$ for the strong coupling case. Colors are the same as in Fig. 8. Snapshots averaged over 500 realizations. Np-lp transition (near $\mu = +6$) proceeds as an avalanche starting from the boundaries, while lp-np transition (near $\mu = -7$) proceeds through intermittent formation and coalescence of lp phase clusters.

Finally, we studied the finite size effects on the position of breathing transitions (indicated by the chemical potential $\mu$ at which each transition happens) by varying the elastic coupling parameter $c_1$ (here in the range from 1 to 13). This is done for two different system sizes, $L = 25$ and $L = 100$, using a free energy barrier of $E_B = 9$ and averaging over 500 realizations of the adsorption–desorption cycle. As shown in Fig. 11, for all values of the elastic coupling except the lowest one ($c_1 = 1$), the adsorption–desorption isotherms form two clear hysteresis loops, whose width is characterized by the difference in the chemical potentials corresponding to the adsorption and desorption transitions. As described above in the case of our "thermodynamic limit" (L=2000), this width increases with coupling. We also observe that the size effect is greater for wider hysteresis loops, i.e. for systems with larger elastic coupling $c_1$. In this case, the larger the system, the wider the adsorption–desorption hysteresis loop. Furthermore, it can be seen that the size effect on the hysteresis loop width can be mostly attributed to a shift of the adsorption branch, while the desorption branch remains unaffected by system size changes (within certain fluctuations).

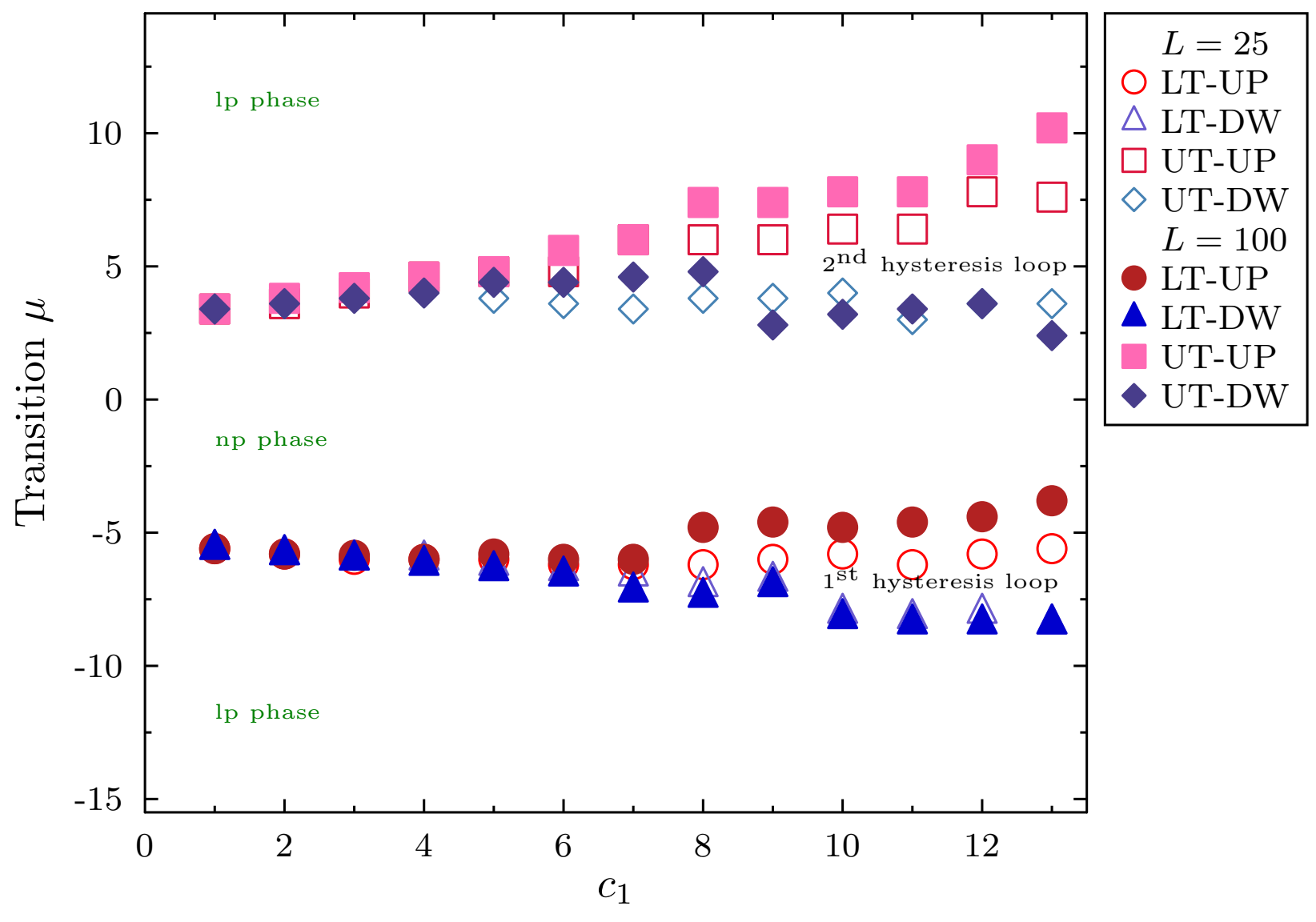

**Fig.11**. The dependence of the transition chemical potentials, $\mu$, on the elastic coupling parameter $c_1$ for two system sizes, $L = 25$ and $L = 100$, at $E_B = 9$. The lower series corresponds to the 1st hysteresis loop at low $\mu$; open (L = 25) and filed (L = 100) circles and triangles show the positions of adsorption and desorption transitions respectively. The upper corresponds the 2nd hysteresis loop at high $\mu$; open (L = 25) and filed (L = 100) squares and diamonds show the positions of adsorption and desorption transitions, respectively. The distance between open and filled symbols represents the width of the hysteresis loop.

## IV. Conclusion

The theoretical description of adsorption-induced structural transformations in flexible metal–organic frameworks so far have been mostly focused on structural, energetic and thermodynamic properties at the microscopic level. Empirical observations have also been discussed, such as the hysteretic nature of these transformations between two metastable phases and the phase co-existence within the transition region. We presented a model linking the microscopic behavior on the level of the unit cell to the dynamics of the adsorption-induced structural transition on the level of the entire crystal, using the archetypal MIL-53 "breathing" framework as a case-study system. We considered the material framework as an elastic three dimensional lattice comprised by the unit cells, and introduced a simple Hamiltonian coupling adsorption within the cells and deformation on the framework level. In doing so, we, first, showed that the constraints on the deformations of the neighboring cells leads to long-range elastic interactions within 3D lattice. These constraints cause a homogeneous phase distribution within 2D layers of cells and a layer-by-layer shear mechanism of the transformation dynamics. As such, 3D framework model reduces to 1D model for a stack of the cell layers. Secondly, we introduced and studied a Monte Carlo simulation model of the crystal dynamics. We showed that this model reproduces the known phenomenology of breathing transitions and investigated the influence of key physical parameters, the free energy barrier of phase transformation, the cell-cell elastic interaction, and the system size, on the system hysteretic behavior and the dynamics of phase nucleation and growth.

In particular, we determined the regions of system parameters, which correspond to reversible and hysteretic transformations, and identified two different dynamic regimes, intermittent nucleation and growth of new phase clusters and avalanche-type phase growth from the crystal boundaries. Our main conclusion regarding the shear layer-by-layer mechanism of phase transformation and the results of MC simulation suggest a possibility of np-lp phase coexistence in the process of breathing transition in one crystal. Such phase coexistence, revealed in experiments [8], may be enhanced and quenched due to various defects in real systems.

**Acknowledgements**

A.V.N. acknowledges NSF ECR "Structural Organic Particulate Systems" and the Région Île-de-France for the support via a Blaise Pascal International Research Chair, administered by the Fondation de l'École normale supérieure.

**References.**

[1] J. Long and O. Yaghi, Chem. Soc. Rev., **38**, 1213–1214 (2009).

[2] G. Férey and C. Serre, Chem. Soc. Rev., **38**, 1380–1399 (2009).

[3] S. Horike, S. Shimomura, and S. Kitagawa, Nat. Chem., **1**, 695–704 (2009).

[4] C. Serre, S. Bourrelly, A. Vimont, N.A. Ramsahye, G. Maurin, P.L. Llewellyn, M. Daturi, Y. Filinchuk, O. Leynaud, P. Barnes, and G. Férey, Adv. Mater. **19**, 2246 (2007).

[5] Y. Liu, J.-H. Her, A. Dailly, A. Ramirez-Cuesta, D.A. Neuman, and C.M. Brown, J. Am. Chem. Soc., **130**, 11813–11818 (2008).

[6] I. Beurroies, M. Boulhout, P.L. Llewellyn, B. Kuchta, G. Férey, C. Serre, and R. Denoyel, Angew. Chem., Int. Ed., **49**, 7526–7529 (2010).

[7] A. Boutin, M.-A. Springuel-Huet, A. Nossov, A. Gédéon, T. Loiseau, C. Volkringer, G. Férey, F.-X. Coudert, and A.H. Fuchs, Angew. Chem. Int. Ed. **48**, 8314–8317 (2009).

[8] C. Serre, S. Bourrelly, A. Vimont, N.A. Ramsahye, G. Maurin, P.L. Llewellyn, M. Daturi, Y. Filinchuk, O. Leynaud, P. Barnes, G. Férey, Adv. Mater., **19**, 2246 (2007).

[9] C. Serre, F. Millange, C. Thouvenot, M. Noguès, G. Marsolier, D. Louër, G. Férey, J. Am. Chem. Soc., **124**, 13519–13526 (2002).

[10] F. Salles, A. Ghou, G. Maurin, R.G. Bell, C. Mellot-Draznieks, and G. Férey, Angew. Chem. Int. Ed., **47**, 8487–8491 (2008).

[11] C. Triguero, F.-X. Coudert, A. Boutin, A. H. Fuchs, A. V. Neimark, J. Phys. Chem. Lett., **2**, 2033–2037 (2011).

[12] A.J.C. Barre de Saint-Venant in C.L.M.H. Navier, Résumé des Leçons sur l'Application de la Mechanique (Dunod, Paris, 1864).

[13] R. C. Albers, R. Ahluwalia, T. Lookman, and A. Saxena, Comp. App. Math., **23**, 345–361 (2004).

[14] M.H. Sadd, *Elasticity.* (Elsevier, 2005) ,pp. 37–41.

[15] M. P. Ariza and M. Ortiz, Arch. Rational Mech. Anal. **178**, 149–226 (2005).

[16] V.V. Bulatov and W. Cai, *Computer simulations of dislocations* (Oxford University Press, 2006), pp. 10–11.

[17] M. Baus and R. Lovett, Phys. Rev. Lett., **65**, 1781 (1990).

[18] M. Baus and R. Lovett, Phys. Rev. A, **44**, 1211 (1991).

[19] J. S. Rowlinson, Phys. Rev. Lett., **67**, 406 (1991).

[20] K. Ø. Rasmussen, T. Lookman, A. Saxena, A. R. Bishop, R. C. Albers, and S. R. Shenoy, Phys. Rev. Lett., **87**, 055704 (2001).

[21] T. Lookman, S.R. Shenoy, K.Ø. Rasmussen, A. Saxena, and A.R. Bishop, Phys. Rev B, **67,** 024114 (2003).

[22] G. F. Smith and R.S. Rivlin, Trans. Am. Math. Soc., **88**, 175 (1958).

[23] K. Bhattacharya and A. Schlömerkemper, ARMA, **196**, 715–751 (2010).Potentials

[24] S.R. Shenoy, T. Lookman, A. Saxena, and A.R. Bishop, Phys. Rev. B, **60**, R12537 (1999).

[25] J. C. Tolédano and P. Tolédano, *The Landau theory of phase transitions* (World Scientific Lecture Notes in Physics, 1987).

[26] I. Müller and P. Villaggio, Arch. Rat. Mech. Anal., **65**, 25–46 (1977).

[27] A. Neimark, F.-X. Coudert, C. Triguero, A. Boutin, A.H. Fuchs, I. Beurroies, and R. Denoyel, Langmuir, **27**, 4734 (2011).

[28] H. C. Kang and W.H. Weinberg., J. Chem. Phys., **90**, 2824 (1989).

[29] A.V. Neimark, F.-X. Coudert, A. Boutin, A.H. Fuchs, J. Phys. Chem. Lett., **1**, 445 (2010).